\documentclass[12pt,letterpaper]{article}
\usepackage{osajnl}

\begin{document}


\title{Ghost imaging with intense fields
from chaotically-seeded parametric downconversion}
\author{Emiliano Puddu and Alessandra Andreoni}
\affiliation{Dipartimento di Fisica e Matematica, Universit\`a degli
Studi dell'Insubria,\\Istituto Nazionale per la Fisica della Materia
(C.N.R-I.N.F.M.), Como, Italy}
\author{Ivo Pietro Degiovanni}
\affiliation{Istituto Nazionale di Ricerca Metrologica, Torino,
Italy}
\author{Maria Bondani}
\affiliation{National Laboratory for Ultrafast and Ultraintense
Optical Science, C.N.R.-I.N.F.M., Como, Italy}
\author{Stefania Castelletto}
\affiliation{Physics School, Melbourne University, Victoria,
Australia}

\begin{abstract}
We present the first experimental demonstration of ghost imaging
realized with intense beams generated by a parametric down
conversion interaction seeded with pseudo-thermal light. As
expected, the real image of the object is reconstructed satisfying
the thin-lens equation. We show that the experimental visibility of
the reconstructed image is in accordance with the theoretically
expected one.
\end{abstract}
\begin{figure}[h]
\begin{center}
\includegraphics[angle=90, width=15cm]{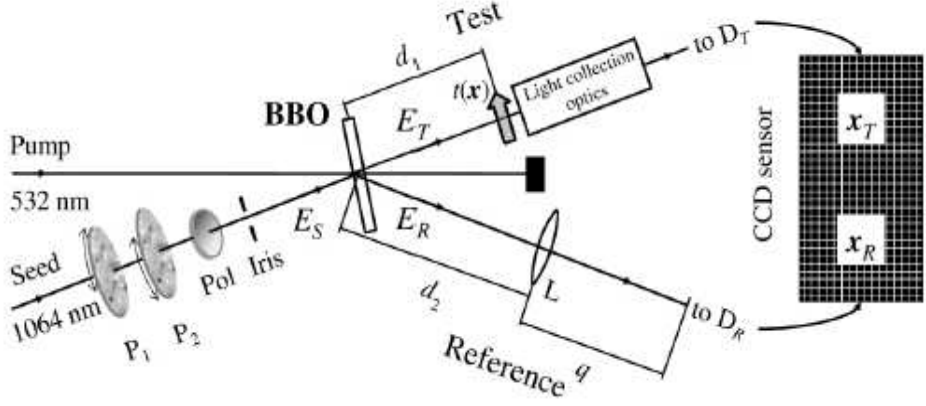}
\end{center}
\caption{Experimental setup: P$_{1,2}$, independently rotating
ground glass plates: Pol, polarizer; $t(\mathbf{x})$, object
transmission function; L, lens; tuning angle 24.7 deg.}
\label{ExpSetup}
\end{figure}
\begin{figure}[h]
\begin{center}
\includegraphics[angle=270, width=15cm]{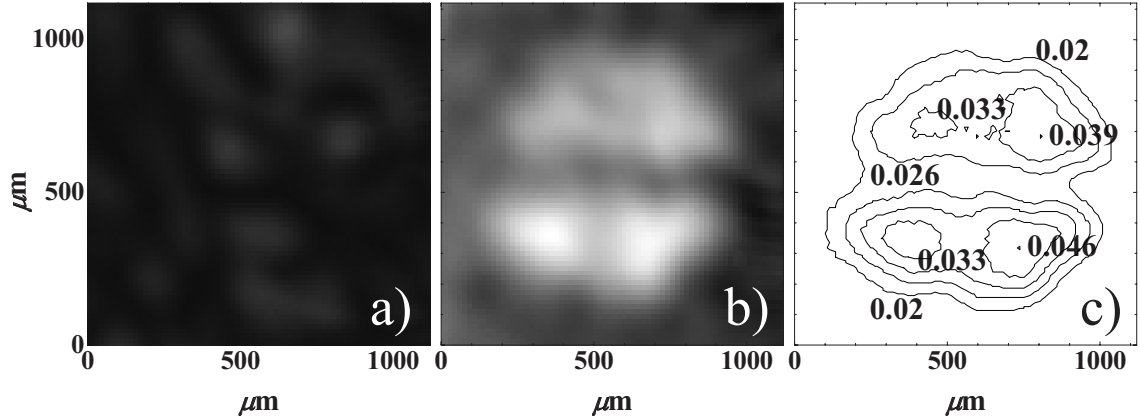}
\end{center}
\caption{a) Single-shot Map recorded in the R-arm; b) ghost image;
c) contour plot of the local visibility. }\label{Results}
\end{figure}

%
%
Ghost imaging capability was initially attributed only to entangled
photon pairs,\cite{Pittman1995} but in more recent works it has been
ascertained also for classically correlated fields, both in the
single photon regime \cite{Valencia2005, Zhang2005} and in the
continuous variable regime.\cite{Ferri2005} The main result of all
these schemes is the non-local imaging of objects both in near- and
far-field. \cite{Dangelo2005} A general ghost-imaging scheme
involves a source of correlated bipartite field and two propagation
arms usually called Test (T) and Reference (R). In the T-arm the
object to be imaged is inserted and a bucket (or a pointlike)
detector measures the light transmitted by the object. The R-arm
contains an optical setup suitable for reconstructing the image of
the object (or its Fourier transform) and a position-sensitive
detector.
\newline\indent
In this Letter we show a ghost imaging experiment performed by using
the intense fields generated by parametric downconversion (PDC)
seeded with multimode chaotic light. As usual in ghost imaging
experiments, the technique adopted for the retrieval of the image
consists in the evaluation of the fourth-order correlation between
the fields at the detection planes.
\newline\indent
In Fig.~\ref{ExpSetup} we show our experimental setup. We injected a
coherent collimated pump beam ($\lambda_P =532$~nm) and a seed beam
($\lambda =1064$~nm) into a crystal of $\beta$-BaB$_2$O$_4$ (BBO,
cut angle 22.8 $\deg$, 10~mm$\times$10~mm$\times$3~mm, Fujian
Castech Crystals) in type-I interaction geometry. Both pump and seed
beams were provided by an amplified Q-switched Nd:YAG laser (GCR-4,
10 Hz repetition rate, 7 ns pulse duration of the fundamental pulse,
Spectra-Physics). The seed field was randomized by passing it
through two independently rotating ground glass plates,
\cite{Arecchi1965} P$_1$ and P$_2$ in order to obtain a
pseudo-thermal statistics. The speckles on the object plane resulted
to be about 320 $\mu$m in diameter as evaluated by spatial
autocorrelation. The object was a hole of 1.6~mm diameter crossed by
a straight wire of 0.5~mm caliber. As the D$_T$ and  D$_R$
detectors, we used a single CCD camera (CA-D1-256T, 16
$\mu$m$\times$16 $\mu$m pixel area, 12 bit resolution, Dalsa). The
CCD sensor, also depicted in Fig.~\ref{ExpSetup}, for one half
recorded the single-shot intensity Maps in the R-arm,
$\mathcal{I}_R( \mathbf{x}_R)$, and for the other half realized the
bucket detector of the T-arm.
%
%
%
In the following, $\mathcal{I}_T$ designates the sum of the contents
of $110\times110$ pixels of D$_T$ in a single shot. The T-arm
includes free propagation over $d_1 = 60$~cm from BBO to object,
transmission function $t(\mathbf{x})$, and light collection optics
in front of $D_T$. The R-arm contains free propagation over  $d_2 =
60$~cm from BBO to lens L ($f = 400$~mm) and from L to the CCD
camera ($q = 60$~cm). Distances $d_1$, $d_2$ and $q$ and focal
length $f$ must satisfy $1/(d_1+ d_2)+1/q=1/f$ to give an imaging
system with magnification factor $M =
q/(d_1+d_2)$.\cite{Pittman1995}
%
\newline\indent
The evolution of the system is described by the unitary operator $S$
generated by the multimode PDC hamiltonian \cite{Mandel} and the
input-output relations that link the output operators
$b_{i,\mathbf{q}}$ to the input operators $a_{i,\mathbf{q}}$ are
\begin{equation}
b_{i,\mathbf{q}}= S^{\dagger} a_{i,\mathbf{q}} S = U_{i,\mathbf{q}}
a_{i,\mathbf{q}} + V_{i,\mathbf{q}}
e^{i\varphi_\mathbf{q}}a^\dagger_{j,-\mathbf{q}}~~~~~~ i = R,T;~i
\neq j, \label{eqTH2}
\end{equation}
where $U_{i,\mathbf{q}}^2-V_{i,\mathbf{q}}^2=1$ and
$\{U,V,\varphi\}_{\mathbf{q}}=\{U,V,\varphi\}_{-\mathbf{q}}$. We
take $a_{R,\mathbf{q}}$ in the vacuum state and $a_{T,\mathbf{q}}$
in the multimode thermal state represented by the density matrix
\begin{equation}
\rho_{T} = \prod^\otimes_{\mathbf{q}}\left\{ \sum_{m=0}^{\infty}
p_{T,{\mathbf{q}}}(m) \left|m\right\rangle\left\langle
m\right|_{T,{\mathbf{q}}} \right\}\; ,
\end{equation}
where $\left|m\right\rangle_{\mathbf{q}}$ is the Fock state with $m$
photons in mode $\mathbf{q}$ and $p_{T,{\mathbf{q}}}(m)=
n_{th,\mathbf{q}}^m/(1 + n_{th,\mathbf{q}})^{m+1}$ is the photon
number distribution on the single mode for $n_{th,\mathbf{q}}$ mean
photon number. The the output state is obtained by exploiting the
Baker-Campbell-Hausdorff formula\cite{Mandel,Traux1985}
%
\begin{eqnarray}
&&\rho_{out}= S(\rho_{T}\otimes \left|0\right\rangle\left\langle 0
\right|_{R})S^{\dagger}\nonumber\\
&&\qquad=\prod^\otimes_{\mathbf{q}}\sum_{n} p_{T,{\mathbf{q}}}(n)
\sum_{n_{1},n_{2}}
F_{\mathbf{q}}(n,n_{1},n_{2})\nonumber\\
&&\qquad\quad \left|n+n_{1}\right\rangle\left\langle
n+n_{2}\right|_{T,{\mathbf{q}}} \otimes\left|n_{1}
\right\rangle\left\langle n_{2}\right|_{T,-{\mathbf{q}}}
\end{eqnarray}
where we have defined
\begin{eqnarray}
&&F_{\mathbf{q}}(n,n_{1},n_{2}) =
\frac{\left(n_{\mathrm{PDC},\mathbf{q}}\right)^\frac{n_{1}+n_{2}}{2}}
{\left(1+n_{\mathrm{PDC},\mathbf{q}}\right)^{n+1+\frac{n_{1}+n_{2}}{2}}}
\nonumber\\ &&\qquad\qquad\qquad\times\sqrt{\frac{(n+
n_{1})!}{n!n_{1}!}} \sqrt{\frac{(n+ n_{2})!}{n!n_{2}!}}
\end{eqnarray}
$n_{\mathrm{PDC},\mathbf{q}}=V_{R,\mathbf{q}}^2$ being the mean
photon number per mode generated by spontaneous PDC. The marginal
distributions on $R$- and $T$-arms are multithermal.
\newline
By direct application of the Peres-Horodecki-Simon
criterion\cite{Peres1996, Horodecki1997, Simon2000} it can be
demonstrated that the state $\rho_{out}$ is inseparable for any
value of $n_{th,\mathbf{q}}$.\cite{manuscript}
\newline\indent
The reconstruction of the image is achieved by the computation of
the correlation function, $G^{(2)}(\mathbf{x}_R)$. This procedure is
equivalent first to evaluate the correlation function between
$I_R(\mathbf{x}_R)$ and $I_T(\mathbf{x}_T)$
\begin{equation}
\mathcal{G}^{(2)}(\mathbf{x}_R, \mathbf{x}_T) = \langle I_R
(\mathbf{x}_R) I_T (\mathbf{x}_T) \rangle - \langle I_R
(\mathbf{x}_R)\rangle \langle I_T (\mathbf{x}_T) \rangle ,
\label{eqTH1}
\end{equation}
where $\langle I_i (\mathbf{x}_i)\rangle = \langle
c_i^\dagger(\mathbf{x}_i) c_i(\mathbf{x}_i)     \rangle $ ($i=R,T$)
is the mean intensity of the {\it i}-th beam and $\langle I_R
(\mathbf{x}_R) I_T (\mathbf{x}_T) \rangle = \langle c_R^\dagger
(\mathbf{x}_R) c_R (\mathbf{x}_R) c_T^\dagger (\mathbf{x}_T) c_T
(\mathbf{x}_T) \rangle$ (where $\langle ... \rangle =
\mathrm{Tr}[... \rho_{T}\otimes \left|0\right\rangle\left\langle 0
\right|_{R}]$), and then to perform the integration
\begin{equation}
G^{(2)}(\mathbf{x}_R) = \int{d\mathbf{x}_T
\mathcal{G}^{(2)}(\mathbf{x}_R,\mathbf{x}_T)} .
\end{equation}
The propagation to $D_R$ and $D_T$ is described by the corresponding
impulse response functions $h_R (\mathbf{x}_{R},\mathbf{x}_{R}')$
and $h_T (\mathbf{x}_{T},\mathbf{x}_{T}')$, whose derivation is
straightforward, \cite{Goodman} and the field operators at the
detection planes become $c_i (\mathbf{x}_i) = \int {\rm d}
\mathbf{x}_i' h_i( \mathbf{x}_i,\mathbf{x}_i') b_i (\mathbf{x}_i')$,
where $b_i (\mathbf{x}) \propto \sum_{\mathbf{q}}
\exp(i\mathbf{q}\cdot \mathbf{x}) b_{i,\mathbf{q}} $. The
factorization rule for $\langle b_R^\dagger (\mathbf{x}_R'') b_R
(\mathbf{x}_R') b_T^\dagger (\mathbf{x}_T'') b_T (\mathbf{x}_T')
\rangle$ in this case is exactly the same as that for spontaneous
PDC\cite{Brambilla2004}, thus in the degenerate case
Eq.~(\ref{eqTH1}) becomes
\begin{eqnarray}
&&\mathcal{G}^{(2)}(\mathbf{x}_R, \mathbf{x}_T) = \nonumber\\
&&\left| \int {\rm d} \mathbf{x}_R' \int {\rm d} \mathbf{x}_T'  h_R
(\mathbf{x}_R, \mathbf{x}_R') h_T (\mathbf{x}_T, \mathbf{x}_T')
\langle b_R (\mathbf{x}_R') b_T(\mathbf{x}_T') \rangle \right|^2 .
\label{eqTH4}
\end{eqnarray}
If the coherence area of the multithermal field is much smaller than
the object, we can proceed as in Ref. \cite{Brambilla2004} and get
%
\begin{eqnarray}
&&G^{(2)}(\mathbf{x}_R)\propto \left| t(-M \mathbf{x}_R)\right|^2
\int {\rm d} \mathbf{x}_T   n_{\mathrm{PDC}} \left(\frac{2 \pi
\mathbf{x}_{T}}{\lambda f}\right)\nonumber\\
&&\times\left[1+ n_{\mathrm{PDC}} \left(\frac{2 \pi
\mathbf{x}_{T}}{\lambda f}\right) \right] \left[1+ n_{th}
\left(\frac{2 \pi \mathbf{x}_{T}}{\lambda f}\right) \right]^2.
\label{eqTH5}
\end{eqnarray}
In our experiment we estimate $n_{\mathrm{PDC}}\simeq 0.5$ and
$n_{th}\simeq 10^{12}$ per spatial mode.  The correlation function
in Eq.~(\ref{eqTH5}) reproduces the object and depends on the
properties of our source through the mean photon numbers of
downconverted fields and of the chaotic seeding field.
\newline\indent
The experimental results are shown in Fig.~\ref{Results}: panel a)
displays a typical sample of chaotic intensity Map $\mathcal{I}_R(
\mathbf{x}_R)$ recorded in single shot, whereas panel b) contains
the map of the correlation coefficients
\begin{eqnarray}
C(\mathbf{x}_R)=\frac{G^{(2)}(\mathbf{x}_R)}{\sqrt{\sigma^2(I_R(\mathbf{x}_R))
\sigma^2(I_T)}}\label{coeffCORR}
\end{eqnarray}
evaluated over an ensemble of 9500 single shot Maps, where
$\sigma^2(I)=\langle I^2\rangle-\langle I\rangle^2$ and both
intensities and variances were evaluated by subtracting the
statistical contributions of the background noise. To measure the
noise we considered the values recorded by the CCD in a
non-illuminated sensor region. Panel b) in Fig.~\ref{Results} shows
that the recovered image has the correct size, as compared to the
original object, being $M = 0.5$. %
%
\newline\indent
To quantify the quality of our technique, we define the local
visibility of the reconstructed image (see Ref.\cite{Gatti2006}) as
\begin{equation}
\mathcal{V}(\mathbf{x}_R)= \frac{G^{(2)}(\mathbf{x}_R)}{\langle
I_R(\mathbf{x}_R) I_T\rangle}= \frac{G^{(2)}(\mathbf{x}_R)}{\langle
I_R(\mathbf{x}_R)\rangle \langle I_T\rangle+ G^{(2)}(\mathbf{x}_R)}
\label{visib}
\end{equation}
and evaluate it for the ghost image in Fig.~\ref{Results} b).
Figure~\ref{Results} c) shows the contour plot of
$\mathcal{V}(\mathbf{x}_R)$ at the marked levels including the
maximum value 0.046.
\begin{figure}[t]
\begin{center}
\includegraphics[angle=0, width=15cm]{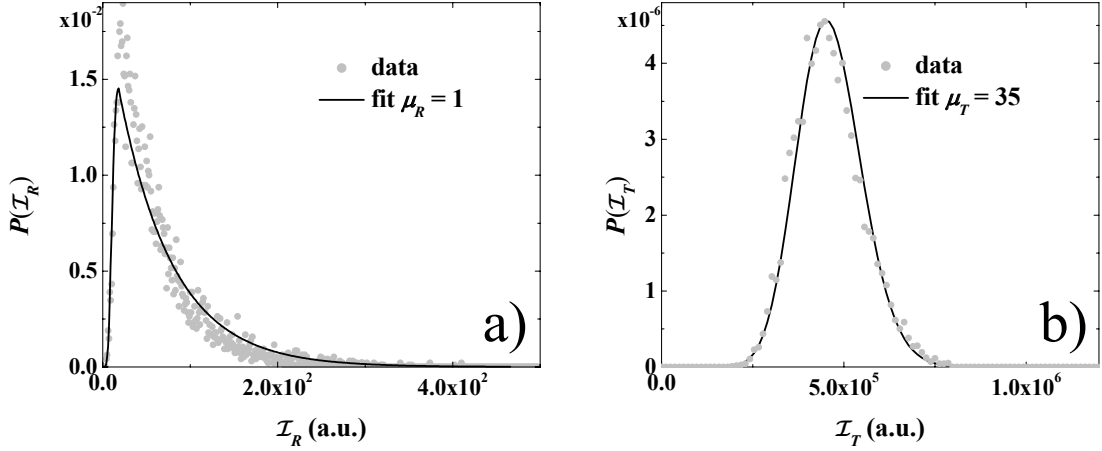}
\end{center}
\caption{a) Statistical distribution of the intensity in a single
pixel of the $D_R$ detector (dots) and multithermal fit giving
$\mu_R=1$ (full line); b) statistical distribution of the intensity
on the bucket detector (dots) and multithermal fit giving $\mu_T=35$
(full line).}\label{Modes}
\end{figure}
\newline\indent
A tradeoff between visibility and resolution of the reconstructed
image has been reported\cite{Gatti2006} that depends on the number
of spatial modes (coherence areas, $A_{C}$) that illuminate an
object of area $A_{O}$. In our case $A_{O}/A_{C}\simeq
1600^2/320^2=25$. This number being rather low, we observe a quite
poor image resolution and a quite high visibility. To confirm our
results we calculate the number of spatial and temporal modes,
involved in the interaction. We start by studying the statistical
distributions of the beam intensities detected either in a single
pixel $\mathbf{x}_R$ or by the bucket detector in the T-arm. In
Fig.~\ref{Modes} we show the experimental distributions along with
the best fitting curves obtained by convolving the theoretical
multithermal distribution\cite{Paleari2004,Mandel}
\begin{equation}
P_\mu(\mathcal{I}) = \frac{e^{-\mu \mathcal{I}/\langle
I\rangle}}{\left(\langle I\rangle/\mu\right)^\mu}
\frac{\mathcal{I}^{\mu-1}}{\left(\mu-1\right)!} \label{multi}\:
\end{equation}
with the experimental one obtained for the background. Here
$\mathcal{I}$ indicates the single-shot values of either
$\mathcal{I}_R(\mathbf{x}_R)$ or $\mathcal{I}_T$, whose mean values
and variances obviously coincide with those of the operators
$I_R(\mathbf{x}_R)$ and $I_T$. From the data in panel a) we obtained
$\mu_R=1$, independently of the position $\mathbf{x}_R$ we tested,
and from the data in panel b), $\mu_T=35$. As the statistical
distribution in Fig.~\ref{Modes} a) reflects the behavior of the
field from shot to shot, we interpret $\mu_R$ as the number of
temporal modes in the field.\cite{Paleari2004,Mandel} On the other
hand, we can interpret $\mu_T$ as the product of the number of
temporal modes times that of the spatial modes in the area covered
by the $110\times110$ pixels of $D_T$. As $\mu_R\simeq 1$ we
conclude that we have a number of spatial modes in the bucket, 35,
greater than the number, 25, of those illuminating the object (see
above): this is to be expected as the bucket integration covered an
area wider than $A_{O}$. From Eq.~(\ref{multi}) we get
$\sigma^2(I_{R,T})= \langle I_{R,T}\rangle^2/\mu_{R,T}$, which
allows linking the visibility of Eq.~(\ref{visib}) to the number of
thermal modes in the incoherent T- and R-beams. By using
Eq.~(\ref{coeffCORR}) we can rewrite Eq.~(\ref{visib}) as
\begin{eqnarray}
\mathcal{V}(\mathbf{x}_R)&=& \frac{C(\mathbf{x}_R)}{\langle
I_R(\mathbf{x}_R)\rangle \langle
I_T\rangle/\sqrt{\sigma^2(I_R(\mathbf{x}_R)) \sigma^2(I_T)}+
C(\mathbf{x}_R)}\nonumber\\
&=&\frac{C(\mathbf{x}_R)}{\sqrt{\mu_R\mu_T}+
C(\mathbf{x}_R)}\label{visib2}\;
\end{eqnarray}
and make an independent estimation of $\sqrt{\mu_R\mu_T}$ as a
function of the experimental values of correlation coefficients and
visibility. By this method we find $\sqrt{\mu_R\mu_T}\simeq 6.8$. As
from the fits in Fig.~\ref{Modes} we found the value $\simeq 5.9$,
the experimental results are self-consistent.
\newline\indent
In conclusion, we have implemented a new source of correlated beams
suitable to perform ghost imaging and, as we expect, ghost
diffraction experiments. Besides the advantage of using detectors
that measure intense light, the major benefit of this source as
compared to those operating on spontaneous PDC is the possibility of
tuning resolution and visibility of the ghost image, for instance by
modifying the spatial coherence properties of the seed beam. In fact
we observed that the spatial-coherence structure of the seed is
preserved on both arms of the seeded PDC emission because all its
spatial components undergo interaction with the pump. The only
limitation could arise from the competition between the speckle
divergence ($< 0.3$~mrad) and the angular bandwidth of the
interaction ($\sim 5$~mrad). Note that the contribution of
spontaneous PDC, which is expected to have a greater angular spread,
is negligible in our experiment.
\newline\indent
We thank M.G.A. Paris (University of Milan) for fruitful
discussions.


\end{document}